\documentstyle[12pt]{article}

\newcommand{\be}{\begin{equation}}
\newcommand{\ee}{\end{equation}}
\newcommand{\dd}{\partial}
\newcommand{\bea}{\begin{eqnarray}}
\newcommand{\eea}{\end{eqnarray}}

\begin{document}
\baselineskip .25in
\newcommand{\numero}{hep-th/9605070, SHEP 96/12}  
 %Enter SHEP preprint number

\newcommand{\titre}{A Novel Symmetry in Sigma models}
\newcommand{\auteura}{Noureddine Mohammedi} 
\newcommand{\place}{Department of Physics\\University of
Southampton\\
Southampton SO17 1BJ \\ U.K. }
\newcommand{\beq}{\begin{equation}}
\newcommand{\eeq}{\end{equation}}

\newcommand{\abstrait}
{A class of non-linear sigma models possessing a new symmetry is
identified. The same symmetry is also present in Chern-Simons
theories. This hints at a possible topological origin for
this class of sigma models. The non-linear sigma models 
obtained by non-Abelian duality are a particular case in this class.}
\begin{titlepage}
\hfill \numero  \\
\vspace{.5in}
\begin{center}
{\large{\bf \titre }}
%\bigskip \\ by
\bigskip \\ \auteura 
\,\,\footnote{e-mail: nouri@hep.phys.soton.ac.uk}
 %\bigskip \\  \auteurb \\
 % \bigskip  and \bigskip \\ \auteurc 
    \bigskip \\ \place \bigskip \\

\vspace{.9 in} 
{\bf Abstract}
\end{center}
\abstrait
 \bigskip \\
\end{titlepage}
\newpage

\section{Introduction}

Non-linear sigma models in two dimensions possess remarkable 
features due to their rich symmetries. The symmetry properties 
a sigma model can have depend very much on the form of the 
metric and the torsion tensors. The most studied 
non-linear sigma model is the Wess-Zumino-Novikov-Witten
(WZNW)
model. This model enjoys an extra symmetry,
generating a current algebra \cite{witty}, precisely when the metric 
and the torsion tensors take the very specific forms
\bea
G_{ij} &=& e^a_ie^b_j\eta_{ab}
=\widetilde{e}^a_i\widetilde{e}_j^b\eta_{ab}
\nonumber\\
H&=&-{1\over 3}{\mathrm{tr}}\left(e\wedge e\wedge e\right)
={1\over 3}{\mathrm{tr}}\left(\widetilde {e}\wedge 
\widetilde{e}\wedge \widetilde{e}\right )\,\,\,,
\eea
where we have introduced the differential forms on the group manifold
\cite{brateen}
\bea
e&=& T_a e^a_i{\mathrm{d}}\phi^i
=g^{-1}{\mathrm{d}}g\nonumber\\
\widetilde{e}&=& T_a \widetilde{e}^a_i{\mathrm{d}}\phi^i
=-{\mathrm{d}}g g^{-1}\,\,\,.
\eea
The torsion tensor $H={1\over 3!}H_{ijk}{\mathrm{d}}\phi^i
\wedge{\mathrm{d}}\phi^j\wedge{\mathrm{d}}\phi^k$ is related to
the antisymmertic tensor field $B={1\over 2!}B_{ij}
{\mathrm{d}}\phi^i\wedge{\mathrm{d}}\phi^j$ by the usual 
relation $H={1\over 2} {\mathrm{d}}B$. 
Here $\eta_{ab}$ is the invariant bi-linear form of the 
Lie algebra generated by $T_a$ and $g$ is a group element
parametrised by $\phi^i$.
\par
In fact one can ask whether there exist other forms
for $G_{ij}$ and $H_{ijk}$ which would lead to 
other forms of current algebras. This question was
answered in refs.\cite{denis,peter,denis1,rajeev,jon} 
where a generalisation 
of the above expressions for the metric and the torsion
were found. The WZNW model is then just a particular
case of this generalisation.
\par
We explore, in this paper, the possibility of 
finding other non-linear sigma models that might have 
further symmetries depending on the forms of 
$G_{ij}$ and $B_{ij}$. Indeed, we identify a class
of sigma models which have a very interesting symmetry.
Remarkably, the same symmetry appears in Chern-Simons
theories in three dimensions. This hints  at a possible 
connection between the two theories.

\section{The new symmetry}

Consider the action for a general bosonic two-dimensional non-linear
sigma model 
\be
S(\varphi)=\int {\mathrm {d}}^2x \sqrt{\gamma}\left(
\gamma^{\mu\nu}G_{ij}\left(\varphi\right)\dd_\mu\varphi^i\dd_\nu\varphi^j
+\widehat{\epsilon}^{\mu\nu}B_{ij}\left(\varphi\right)
\dd_\mu\varphi^i\dd_\nu\varphi^j \right)\,\,\,.
\ee
In this equation $\gamma_{\mu\nu}$ is the metric on the two-dimensional
world sheet, $\gamma$ is its determinant and 
$\widehat\epsilon^{\mu\nu}={1\over\sqrt\gamma}
\epsilon^{\mu\nu}$ is 
the alternating tensor. This action can be written as
\be
S(\varphi)=\int {\mathrm {d}}^2x \sqrt{\gamma}\left(
\widehat{\epsilon}^{\mu\nu}\eta_{ij}
A^i_\mu\dd_\nu\varphi^j \right)\,\,\,,
\label{2}
\ee
where we have introduced the gauge field-like quantity $A^i_\mu$ 
\bea
A^i_\mu &=& R^{ij}_{\mu\nu}
\eta_{jk}\widehat {\epsilon}^{\nu\alpha}
\dd_\alpha\varphi^k 
\nonumber\\
R^{ij}_{\mu\nu} &=& \eta^{ik}\eta^{jl} 
\left(\gamma_{\mu\nu}G_{kl} + \widehat{\epsilon}_{\mu\nu}
B_{kl}\right)
\label{3}
\eea
with $\eta_{ij}$ a symmetric field-independent 
metric whose inverse is $\eta^{ij}$. Suppose now that $\eta_{ij}$ is
the invariant bi-linear form of a Lie algebra whose structure
constants we denote by $f^i_{\,\,jk}$ (which means that
$\eta_{ij}f^j_{\,\,kl}+\eta_{kj}f^j_{\,\,il}=0$).
\par
We would like to investigate under which conditions the
action (\ref{2}) has a symmety of the form
\be 
\delta\varphi^i= f^i_{\,\,jk}\xi^j F^k_{\mu\nu}
\widehat{\epsilon}^{\mu\nu}\,\,\,,
\label{4}
\ee
where $\xi^j(x)$ is the infinitesimal gauge parameter and 
$F^i_{\mu\nu}=\dd_\mu A^i_\nu -\dd_\nu A^i_\mu
+ f^i_{\,\,jk}A^j_\mu A^k_\nu$ is the field strength of
the gauge field $A_\mu^i$ as given by (\ref{3}). 
The
transformation is suggested by the form of the 
action (\ref{2}). The same kind of symmetry
was identified in the context of non-Abelian duality
in sigma models \cite{me1} and in non-Abelian 
gauge theories \cite{me2}
\par
We found that the action is invariant, up to a total
derivative, provided that the metric $G_{ij}$ and the antisymmetric
tensor $B_{ij}$ satisfy
\be
\dd_k R^{ij}_{\mu\nu}
=\eta_{kl}f^l_{\,\,mn}
R^{im}_{\mu\alpha}R^{jn}_{\nu\beta}
\widehat{\epsilon}^{\alpha\beta}\,\,\,.
\label{6}
\ee
This condition can of course be  expressed explicitly as two
separate conditions on $G_{ij}$ and $B_{ij}$
\bea
\dd_k G_{ij} &=& f^l_{\,\,km}\eta^{mn}\left(
G_{li}B_{nj} -G_{nj}B_{li}\right)
\nonumber\\
\dd_k B_{ij} &=& -f^l_{\,\,km}\eta^{mn}\left(
G_{li}G_{nj} + B_{li}B_{nj}\right)
\eea
This shows the special geometry of this class of sigma models.
In particular, the Riemann tensor and the torsion will be 
given in closed forms as products of $G_{ij}$,  $G^{ij}$ and $B_{ij}$. 
\par
Under these conditions, the equations of motion of the non-linear
sigma model lead to
\be
\widehat{\epsilon}^{\mu\nu}F^i_{\mu\nu} = 0\,\,\,.
\label{99}
\ee
Therefore the above transformation vanishes on-shell.
As seen later, this equation can also be thought of as deriving from 
a Chern-Simons theory.
\par
It is straightforward to find the unique  
solution to the symmetry invariance condition in (\ref{6}). 
In order to
do this, we denote by $\widetilde{R}^{\mu\nu}_{ij}$ the
inverse of $R^{ij}_{\mu\nu}$ (that is,
$R^{ij}_{\mu\nu}\widetilde{R}
^{\nu\alpha}_{jk} =\delta_\mu^\alpha
\delta^i_k$). Equation (\ref{6}) is then cast into the first
order differential equation
\be
\dd_k \widetilde{R}^{\mu\nu}_{ij}
=-\eta_{kl}f^l_{\,\,ij}\widehat{\epsilon}^{\mu\nu}\,\,\,
\ee
whose general solution is given by 
\be
\widetilde{R}^{\mu\nu}_{ij} =
-\left[N^{\mu\nu}_{ij} + \widehat{\epsilon}^{\mu\nu}
\eta_{kl}f^l_{\,\,ij}\varphi^k\right]
\,\,\,,
\ee
where $N^{\mu\nu}_{ij} =N^{\nu\mu}_{ji}$ is any
field-independent matrix, and in general 
$N^{ij}_{\mu\nu}=\gamma^{\mu\nu} A_{ij} +
{\widehat{\epsilon}}^{\mu\nu} C_{ij}$.
\par
The action can be cast into a form which is familiar 
in the context of
non-Abelian duality \cite{a1,a2,a3,a4,a5,a6,a7,a8,a9,a10}. 
By extracting $\dd_\mu \varphi ^i$ 
from (\ref{3}) and eliminating it in (\ref{2}), one finds,
after some straightforward manipulations, the following action
\be
S(\varphi, A)=\int {\mathrm {d}}^2x \sqrt{\gamma}
\left(
N^{\mu\nu}_{ij} A_\mu^iA_\nu^j +
\widehat{\epsilon}^{\mu\nu}\eta_{ij}
F^i_{\mu\nu}\varphi^j \right)\,\,\,,
\label{dual1}
\ee
where we have ignored a total derivative.
If one treats $A_\mu^i$ and $\varphi^i$ as independent variables
then the equations of motion for $A_\mu^i$ are precisely those
in (\ref{3}). Indeed, this action is obtained (when $A_\mu^i$ 
and $\varphi^i$ are independent) by performing a non-Abelian 
duality transformation on the following action
\be
S(g)=\int {\mathrm {d}}^2x \sqrt{\gamma}
N^{\mu\nu}_{ij}\eta^{ik}\eta^{jl}
{\mathrm{tr}}\left(T_k g^{-1}\dd_\mu g\right)
{\mathrm{tr}}\left(T_l g^{-1}\dd_\nu g\right)\,\,\,,
\label{13}
\ee
where $T_i$ are the generators of  the Lie algebra 
$\left[T_i\,,\,T_j\right] = f^k_{\,\,ij}T_k$,
$g$ is a Lie group element and ${\mathrm{tr}}$ is 
the invariant bi-linear form ${\mathrm{tr}}\left(XY\right)
=\eta_{ij}X^iY^j$. 
This action is invariant 
under the global transformation $g\longrightarrow  hg$. The 
non-Abelian dual theory is obtained by gauging this symmetry
and at the same time restricting the
gauge  field strength to vanish \cite{verlinde}.
We therefore obtain the action
\be
S\left(g,\varphi,A\right)=\int {\mathrm {d}}^2x \sqrt{\gamma}
\left[N^{\mu\nu}_{ij}\eta^{ik}\eta^{jl}
{\mathrm{tr}}\left(T_k g^{-1} D_\mu g\right)
{\mathrm{tr}}\left(T_l g^{-1} D_\nu g\right) +
\widehat{\epsilon}^{\mu\nu}{\mathrm{tr}}\left(
\varphi F_{\mu\nu}\right)\right]\,\,\,.
\ee
The covariant derivative is $D_\mu g= \dd_\mu g
+ A_\mu g$
with $A_\mu \longrightarrow h^{-1}A_\mu h -
\dd_\mu h h^{-1}$,
the gauge field is $A_\mu=T_i A^i_\mu$ and the 
Lagrange multiplier is $\varphi=T_i\varphi^i$ with
$\varphi \longrightarrow h\varphi h^{-1}$.
The gauge invariance allows us to choose a gauge such
that $g=1$. In this gauge $S\left(g,\varphi,A\right)$
reproduces precisely $S(g,A)$ as given by (\ref{dual1}).
\par
The dual of the principal chiral model is a special case
of this construction \cite{jev,arkady,cosmas}. 
The dual of the chiral model is obtained when $A_{ij}=
\eta_{ij}$ and $C_{ij}=0$. 
%In this case, however, the action in (\ref{dual1}),
%in addition of being dual to (\ref{13}), is also dual to the 
%chiral model
%\be
%S\left(g\right)=\int {\mathrm {d}}^2x \sqrt{\gamma}
%\gamma^{\mu\nu}
%{\mathrm{tr}}\left[\left\(g^{-1} \dd_\mu g\right)
%\left(g^{-1} \dd_\nu g\right)\right]\,\,\,.
%\ee
%The chiral model is invariant under the global symmetry 
%$g\longrightarrow hgk$ (instead of the $g\longrightarrow 
%hg$ for the model (\ref{13})). The dual of the chiral model
%is obtained by gauging the left or right symmetry and 
%introducing the Lagrange multiplier term in the gauge 
%$g=1$. Hence the model (\ref{dual1}) is dual to two 
%theories when $N_{ij}^{\mu\nu}=\gamma^{\mu\nu}\eta_{ij}$.

\section{Generalisation}
\par
Another important case is obtained by spliting the field 
$\varphi^i$ 
of the previous non-linear sigma model as 
$\varphi^i=(X^a,Y^A)$ and restricting the transformation
to the fields $X^a$ only. 
In this case, the non-linear sigma model action takes 
the form 
\bea
S(X,Y)&=&\int {\mathrm {d}}^2x \sqrt{\gamma}\left(
\widehat{\epsilon}^{\mu\nu}\eta_{ab}
A^a_\mu\dd_\nu X^b 
+\widehat{\epsilon}^{\mu\nu}
\widehat{\epsilon}^{\alpha\beta}
M^{aA}_{\mu\alpha}\eta_{ab}L_{AB}\dd_\beta Y^B\dd_\nu X^b
\right.\nonumber\\
&+& \left. R^{\mu\nu}_{AB}\dd_ \mu Y^A \dd_\nu Y^B
\right)\,\,\,,
\label{10}
\eea
The gauge field $A^a_\mu$ involves both $X^a$ and $Y^A$ through
\be
A^a_\mu = R^{ab}_{\mu\nu}\widehat{\epsilon}^{\nu\alpha}
\eta_{bc}\dd_\alpha X^c + R^{aA}_{\mu\nu}
\widehat{\epsilon}^{\nu\alpha}L_{AB} \dd_\alpha Y^B\,\,\,.
\label{11}
\ee
The different quantities introduced here can depend on both
$X^a$ and $Y^A$ and are such that
\bea
R^{ab}_{\mu\nu} &=& \eta^{ac}\eta^{bd}
\left[\gamma_{\mu\nu} G_{cd} + \widehat{\epsilon}
_{\mu\nu} B_{cd}\right]\nonumber\\
R_{AB}^{\mu\nu} &=& 
\left[\gamma^{\mu\nu} G_{AB} + \widehat{\epsilon}
^{\mu\nu} B_{AB}\right]\nonumber\\
R^{aA}_{\mu\nu} + M^{aA}_{\mu\nu}
&=& 2\eta^{ab}L^{AB}
\left[\gamma_{\mu\nu} G_{bB} + \widehat{\epsilon}
_{\mu\nu} B_{bB}\right]\,\,\,.
\eea
The symmetric matrices $\eta_{ab}$ and $L_{AB}$ are field-independent
and their inverses are, respectively, $\eta^{ab}$ and $L^{AB}$.
We then suppose that  $\eta_{ab}$ is a bi-linear invariant form
of a Lie algebra with structure constants $f^a_{\,\,bc}$. 
\par
Let us find the conditions under which the sigma model (\ref{10})
is invariant under
\be
\delta X^a = f^a_{\,\,bc}\xi^b F^c_{\mu\nu}
\widehat{\epsilon}^{\mu\nu}\,\,\,\,\,\,,\,\,\,\,\,
\delta Y^A =0\,\,\,.
\ee
We find that the action remains invariant, up to a total derivative,
when the following conditions are fulfilled
\bea
M_{\mu\nu}^{aA} &=& R^{aA}_{\mu\nu}\nonumber\\
\dd_c R^{ad}_{\mu\alpha} &=& 
\eta_{cb}f^b_{\,\,er}R^{ae}_{\mu\sigma} R^{dr}_{\alpha\tau}
\widehat{\epsilon}^{\sigma\tau}
\nonumber\\
\dd_c R^{aA}_{\mu\alpha} &=& 
\eta_{cb}f^b_{\,\,er}R^{ae}_{\mu\sigma} R^{rA}_{\tau\alpha}
\widehat{\epsilon}^{\sigma\tau}
\nonumber\\
\dd_c R_{AB}^{\mu\alpha} &=& 
\eta_{cb}f^b_{\,\,er}
L_{AE}L_{BD}
R^{eE}_{\alpha\sigma} R^{rD}_{\tau\beta}
\widehat{\epsilon}^{\sigma\mu}\widehat{\epsilon}^{\beta\nu}
\widehat{\epsilon}^{\alpha\tau}
\,\,\,.
\eea
Notice that this set of equations cannot be obtained from
(\ref{6}) by simply spliting the field $\varphi^i$ as $(X^a, Y^A)$.
The second equation of this set has the unique solution
for the inverse of $R^{ab}_{\mu\nu}$, namely 
$\widetilde{R}_{ab}^{\mu\nu}$, given by
\be
\widetilde{R}_{ab}^{\mu\nu}=-\left[
N^{\mu\nu}_{ab}\left(Y\right)
+ \widehat{\epsilon}^{\mu\nu}\eta_{cd}f^c_{\,\,ab}
X^d \right]
\,\,\,
\ee
The general solution of the remaining 
last two equations is provided by
\bea
R^{aA}_{\mu\nu} &=& R^{ab}_{\mu\alpha} 
W^{\alpha A}_{\nu b}\left(Y\right)
\nonumber\\
R^{\mu\nu}_{AB}&=& 
L_{AE}L_{BD}
R^{ab}_{\tau\rho}
W^{\tau E}_{\sigma a}
W^{\rho D}_{\beta b} 
\widehat{\epsilon}^{\mu\sigma}\widehat{\epsilon}^{\nu\beta}
+ T^{\mu\nu}_{AB}\left(Y\right)
\,\,\,
\eea
with $N^{\mu\nu}_{ab}$, $W^{\mu A}_{\nu a}$
and $T^{\mu\nu}_{AB}$ any arbitrary functions
which depend on the field $Y^A$ only.
\par
Subject to these conditions, the variation with
respect to $X^a$ of our action leads to the equations
of motion $\widehat{\epsilon}^{\mu\nu}F_{\mu\nu}^a=0$,
where $F^a_{\mu\nu}$ is constructed from 
$A^a_\mu$ in (\ref{11}).
\par
Similarly, by extracting $\dd_\mu X^a$ 
from (\ref{11}) and substituting in (\ref{10}),
we find  (up to a total derivative)
\bea
S\left(Y,X,A\right)&=&
\int {\mathrm {d}}^2x \sqrt{\gamma}\left[
T^{\mu\nu}_{AB}\left(Y\right)\dd_\mu Y^A \dd_\nu Y^B +
N^{\mu\nu}_{ab}\left(Y\right)A^a_\mu A^b_\nu
\right.\nonumber\\
&+& \left. 2W^{\mu A}_{\alpha a}\left(Y\right)
L_{AB}
\widehat{\epsilon}^{\alpha\nu}
A^a_\mu \dd_\nu Y^B
+\widehat{\epsilon}^{\mu\nu}\eta_{ab} X^a F^b_{\mu\nu}\right]
\,\,\,.
\label{17}
\eea
Again, the equations of motion for 
$A_\mu^a$, if considered as an independent
field, are precisely those in (\ref{11}).
The symmetry $\delta X^a =
f^a_{\,\,bc}\xi^b F^c_{\mu\nu}\widehat{\epsilon}^{\mu\nu}$
is transparent in this case.
\par
The same procedure can be applied here to find
the non-Abelian dual theory. Consider now the action
\bea
S\left(Y,g,A\right)&=&
\int {\mathrm {d}}^2x \sqrt{\gamma}\left[
T^{\mu\nu}_{AB}\left(Y\right)\dd_\mu Y^A \dd_\nu Y^B +
N^{\mu\nu}_{ab}\left(Y\right)
\eta^{ac}\eta^{bd}
{\mathrm{tr}}\left(T_cg^{-1}\dd_\mu g\right)
{\mathrm{tr}}\left(T_dg^{-1}\dd_\nu g\right)
\right.\nonumber\\
&+& \left. 2W^{\mu A}_{\alpha a}\left(Y\right)
L_{AB}\eta^{ab}
\widehat{\epsilon}^{\alpha\nu}
{\mathrm{tr}}\left(T_bg^{-1}\dd_\mu g\right)
\dd_\nu Y^B\right]
\eea
which is invariant under the left symmetry
$g\longrightarrow hg$. This symmetry can be gauged
by the replacement $\dd_\mu g\longrightarrow 
D_\mu g=\dd_\mu g + A_\mu g$. The dual theory is obtained 
when the Lagrange multiplier
term $\int {\mathrm {d}}^2x \sqrt{\gamma}\widehat{\epsilon}^{\mu\nu}
{\mathrm{tr}}\left(XF_{\mu\nu}\right)$ is added.
Choosing then a gauge such that $g=1$ yields the action
in (\ref{17}).
\par
Notice that in the above model (\ref{17}) we have not assumed any 
transformation for the fields $Y^A$. In fact these fields could 
transform when $T_{AB}^{\mu\nu}$, $N^{\mu\nu}_{ab}$
and $W^{\mu A}_{\nu a}$ are restricted
to satisfy certain conditions as shown below. 
It is found that the theory
in (\ref{17}), when $A^a_\mu$ is treated as an independent field,
has the infinitesimal local gauge symmetry
\bea
\delta Y^A &=& \lambda^aK^A_a\left(Y\right)\nonumber\\
\delta A^a_\mu &=& -\dd_\mu \lambda^a + f^a_{\,\,bc}
\lambda^b A^c_\mu
\label{26}
\eea
provided that the two quantities $T^{\mu\nu}_{AB}$ and 
$K^A_a$ satisfy 
\bea
\dd_E T^{\mu\nu}_{AB} K^E_a + T^{\mu\nu}_{EB}\dd_A K^E_a
+ T^{\mu\nu}_{AE}\dd_B K^E_a&=& {\widehat{\epsilon}}^{\mu\nu}
\left(\dd_A V_{Ba}-\dd_BV_{Aa}\right)
\nonumber\\
K^A_a\dd_A K^B_b - K^A_b\dd_AK^B_a &=& -f^c_{\,\,ab}K^B_c\,\,\,.
\eea
The second equation merely expresses the fact that
the differential operators $K_a=-K^A_a{\dd\over \dd Y^A}$
form  a representation of the Lie algebra defined by
$\eta_{ab}$ and $f^c_{\,\,ab}$. The first equation 
defines the new quantity $V_{Aa}$ which is required to
satisfy
\bea
\dd_D V_{Ab} K^D_c + \dd_A V_{Dc} K^D_b - \dd_D V_{Ac} K^D_b
+ V_{Db}\dd_AK^D_c &=& -f^d_{\,\,cb}V_{Ad}
\nonumber\\
V_{Aa}K^A_b + V_{Ab}K^A_a&=&0\,\,\,\,.
\eea
The remaining two quantities $N^{\mu\nu}_{ab}$
and $W^{\mu A}_{\nu a}$ are then given by
\bea
N^{\mu\nu}_{ab}&=& T^{\mu\nu}_{AB}K^A_aK^B_b
+ {\widehat{\epsilon}}^{\mu\nu} V_{Ab}K^A_a
\nonumber\\
W^{\mu A}_{\nu a} &=& -L^{AE}\left(
\widehat{\epsilon}_{\alpha\nu}
T^{\alpha\mu}_{EB} K^B_a + \delta^\mu_\nu V_{Ea}\right)\,\,\,.
\label{24}
\eea
By writing 
$T^{\mu\nu}_{AB}= \gamma^{\mu\nu} 
G_{AB}(Y) + \widehat{\epsilon}^{\mu\nu}B_{AB}(Y)$,
the equations (\ref{26})--(\ref{24}) are precisely 
the equations needed to gauge the isometries of a general
sigma model with metric $G_{AB}$ and antisymmetric tensor
$B_{AB}$ \cite{ian,chris}. 
Hence the non-linear sigma model obtained 
through  a non-Abelian duality procedure is a particular case of 
this general construction.

\section{Conclusions}

It is worth mentioning that a symmetry similar to the 
one identified for the sigma model exists in Chern-Simons
theory. To see this,   
consider a Chern-Simons theory for some gauge group
$\mathcal{G}$
\be
I\left(A\right) =
\int{\mathrm{d}}^3x\epsilon^{ijk}\left[
{\mathrm{tr}}\left(A_iF_{jk}\right) -
{2\over 3}{\mathrm{tr}}\left(A_iA_jA_k\right)\right]
\,\,\,
\ee
where $i,j,\dots =1,2,3$ and $\epsilon^{123}=1$. Let also
$\mu,\nu,\dots =1,2$ and $\epsilon^{\mu\nu}$ the corresponding alternating
tensor, with $\epsilon^{12}=1$. 
By spliting the three-dimensional indices, the Chern-Simons
action can be written as
\be
I\left(A\right) =
2\int{\mathrm{d}}^3x\epsilon^{\mu\nu}\left[
{\mathrm{tr}}\left(A_3F_{\mu\nu}\right) -
{\mathrm{tr}}\left(A_\mu\dd_3 A_\nu\right)-
{\mathrm{tr}}\left(\dd_\mu\left(A_\nu A_3\right)\right)
\right]
\,\,\,
\ee
It is then clear, if we drop the total divergence term, 
that the Chern-Simons theory has a further
symmetry given by
\be
A_3\longrightarrow A_3 + \epsilon^{\mu\nu}
\left[\xi\,,\, F_{\mu\nu}\right]\,\,\,,
\ee
where $\xi$ is a local Lie algebra-valued function. This symmetry 
is of the form (\ref{4}). Furthermore, varying the Chern-Simons
action with respect to $A_3$ leads to an equation similar to
(\ref{99}). This hints at a deep connection between Chern-Simons
theory and the class of sigma models we identified as
having the new symmetry. We speculate that a non trivial 
compactification to two dimensions of the Chern-Simons theory
would lead to our sigma models.
\par
As mentioned earlier, the new symmetry vanishes on-shell. Therefore,
at the classical level this symmetry has no effects. 
We expect, however, that this symmetry would play a crucial role
at the quantum level.  
We will report elsewhere on the work in progress
regarding the quantisation of these models. The methods 
designed for the quantisation of Chern-Simons theories 
are essential to this investigation \cite{witty1}.
\par
\par
{\bf{Acknowledgements:}} I would like to thank Peter Forg\'acs, 
Peter Hodges and Cosmas Zackos for discussions and correspondence.

\end{document}